\documentclass[journal]{IEEEtran}
\usepackage{comment} 
\usepackage{amsmath}
\usepackage[pdftex]{graphicx}   
\usepackage{lipsum}
\usepackage{comment}
\usepackage[autostyle]{csquotes}
\usepackage{cite}

\usepackage[caption=false]{subfig}
\usepackage{array}
\usepackage{multirow}

\newcommand\MyBox[2]{
  \fbox{\lower0.75cm
    \vbox to 1.7cm{\vfil
      \hbox to 1.7cm{\hfil\parbox{1.4cm}{#1\\#2}\hfil}
      \vfil}%
  }%
}

\ifCLASSINFOpdf
\else
\fi

\begin{document}

%
\title{Robust Oil-spill Forensics and Petroleum Source Differentiation using Quantized Peak Topography Maps}

\author{
    \IEEEauthorblockN{Hamidreza Ghasemi Damavandi\IEEEauthorrefmark{1}, Ananya Sen Gupta\IEEEauthorrefmark{1}, Guadalupe Canahuate\IEEEauthorrefmark{1}, Christopher Reddy\IEEEauthorrefmark{2},Robert Nelson\IEEEauthorrefmark{2}}
    \IEEEauthorblockA{\IEEEauthorrefmark{1}Electrical and Computer Engineering, University of Iowa
    \\\{hamidreza-ghasemidamavandi,ananya-sengupta,guadalupe-canahuate\}@uiowa.edu}
    \IEEEauthorblockA{\IEEEauthorrefmark{2}Department of Marine Chemistry and Geochemistry, Woods Hole, MA
    \\\{creddy, rnelson\}@whoi.edu}
}
\maketitle

\begin{abstract}

Identification and classification of environmental forensics, with the petroleum forensics as the main application, requires an effective technology or method to distinguish between the closely located forensics as they share many main biomarkers. Two-dimensional gas chromatography is one of these technologies with which a petroleum forensic is separated into its chemical compounds, resulting in a three-dimensional image, $GC \times GC$ image. Therefore, distinguishing between two petroleum forensics is equivalent to the comparison between their corresponding $GC \times GC$ images. In this paper, we present a technique, called Quantized Peak Topography Map (QPTM), which results in a better separation between the $GC \times GC$ images. We validate our proposed method on a model dataset, consisting of thirty-four $GC \times GC$ images, extracted from the different parts of the world.

\end{abstract}


\IEEEpeerreviewmaketitle

\section{Introduction}
In this work, we introduce a robust signal processing approach for petroleum fingerprinting, in particular, source forensics in the aftermath of a major oil spill such as the \emph{Deepwater Horizon} disaster, Gulf of Mexico, April 2010. Petroleum fingerprinting has played a crucial role in apportioning the environmental impact of major oil spills and industrial leaks. Furthermore, robust source differentiation between closely spaced oil reservoirs enable determination of underground connectivity of petroleum reservoirs. Such source-specific knowledge can significantly reduce the environmental, financial, and safety risks of multiple drilling points in off-shore rigs. Therefore, robust source fingerprinting for closely spaced petroleum sources is of paramount importance to environmental health, public safety and resource management for the petroleum industry. 
Petroleum identification based upon a pixel-driven compression technique was first introduced in 
\cite{oceans} 
where a representation of the GC$\times$GC Chromatography image called $\rho_{\tau}$-map was applied for the classification of different GC$\times$GC images using their compressed versions. In \cite{oceans}
 the compressed version of a GC$\times$GC image of a petroleum source from one specific geographical region is achieved by identifying the portion of the image which is common among a bunch of sources from that region. In this scenario, in case there is a noise in one the sources, it could affect the whole representative compressed version of the region. 
 In this paper a compressed version of each GC$\times$GC image is achieved using Symbolic Aggregate ApproXimation (SAX) method and  just based upon its own image and hence a robust technique is presented.

We present here a signal processing perspective to this challenging environmental application that combines recent research in peak topography mapping with well-known techniques in high-volume data quantization. The key innovation developed in this article is (Q)uantized (P)eak (T)opography (M)ap: a scaled quantized topography mapping technique that achieves two important signal processing objectives:
\begin{itemize}
\item[(i)] Peak Topography Mapping: We significantly extend the seed idea of peak topography maps, introduced in \cite{patent,asilomar} to include a wider range of peaks, algorithmically isolated from large-scale chromatographic data. Novel extensions include disentangling low signal-to-noise (SNR) ratio peaks otherwise lost against baseline noise and other perturbations typical in chromatographic instruments.
\item[(ii)] Quantization of peak information based on prioritization of peaks based on their local SNR and relevance to source signature within the joint peak profile of petroleum hydrocarbons. This step is critical in robust pattern recognition of petroleum source signatures, which may span hundreds of well-known (target) and unknown (non-target) biomarkers.
\end{itemize}

In synopsis, the QPTM technique proposed here enables efficient source differentiation of petroleum sources, with broader applicability to a wide range of peak profile data, typically generated in gas chromatographic and mass spectrometric instruments. Thus, from a signal processing perspective, the proposed technique is applicable far beyond the application focus, i.e., petroleum forensics, to most chemical interpretation of instrumental raw signals. Typical applications that may benefit from this signal processing method include, but are not limited to, air pollutant studies (.g. fingerprinting key pollutants sampled in urban air), water sustainability investigations (e.g. isolating key toxins in ground and surface water and fingerprinting them to local industrial agents such as a factory or common household cleaning products), among many others.

We have chosen to focus on petroleum forensics as the target application in this work due to its well-known relevance to environmental studies, public health and national economy.  The rest of this paper is organized as follows. 
Section~\ref{related} provides background and related work in signal processing and high-volume data informatics. Sections~\ref{ps} and~\ref{ta} describe the mathematical formulation of the problem and the proposed approach. Sections~\ref{result} and~\ref{discussion} present the experimental results and related discussions. Conclusions are presented in Section~\ref{conclusion}.

\section{Background and Related Work}
\label{related}

A common analytical approach to petroleum forensics is to separate an oil sample into an information-rich distribution of constituent compounds, which span across an intricate network of hundreds, sometimes thousands, of hydrocarbons. In particular, from a pattern recognition perspective, most petroleum forensic techniques focus on the hydrocarbon "biomarker" groups, hopanes and steranes \cite{aeppeli2014}, which are well-known to withstand environmental degradation. In practical terms, this means that oil sampled from grass blades in Louisiana coastline share very similar biomarker distribution with oil sampled from the Macondo well, source of the \emph{Deepwater Horizon} spill. Therefore, robust source fingerprinting of an oil sampled hundreds of miles away from a spill site reduces to accurate and efficient matching of the biomarker peak distribution between sampled crude oil and oil taken from the spill site. 
\subsection{Two-dimensional Gas Chromatography (GC $\times$ GC)} 
\label{gcxgcSection}

\begin{figure*}[h]
 \centering
 \includegraphics[scale=1]{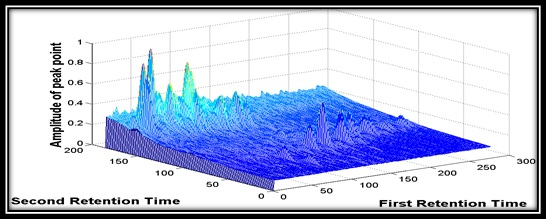}
 \caption{GC$\times$ GC image of a petroleum sample from the Macondo well in the Gulf of Mexico.  }
 \label{gcxgc}
 \end{figure*}

Two-dimensional Gas Chromatography (GC $\times$ GC) is a technology used to separate any petroleum forensic into its constituent components. A petroleum source is injected into a GC$\times$GC system which consists of two stages (columns) where each of the components interact differently with these columns and will pass these columns in different times (retention time). Note that, we have two stages for a better resolution as those components which have not been separated in the first stage can get separated in the second stage. Hence, for each component we will have two retention times corresponding to the time they passed the first and the second columns and also its concentration. The resulting pattern would be a three-dimensional image where the first and second dimensions are the first and second retention times and the third dimension is the concentration of the corresponding component or its peak point (Figure \ref{gcxgc}).

\subsection{Mathematical model of biomarker distribution and related challenges}
Section \ref{gcxgcSection} provides a basic description of two-dimensional gas chromatography (GC$\times$GC), the analytical technology that performs these separation of complex mixtures such as crude oil into constituent compounds. The outcome of GC$\times$GC separations are high-resolution images of hydrocarbon peaks.  Mathematically, the peak distribution of petroleum hydrocarbons in a crude oil sample can be modeled as:
\begin{equation}
\label{model}
I(x,y) = \sum_{i=1}^M \sum_{j=1}^N p_{ij} \psi_{ij}(x-x_i,y-y_j),
\end{equation}
where $I(x,y)$ denotes the pixel intensity at the pixel location $(x,y)$ of the GC$\times$GC image, $(x_i,y_j)$ denotes the peak location for the $j^{th}$ peak along the $i^{th}$ column of the GC$\times$GC image, and $p_{ij}$ denotes the peak height and  $\psi_{ij}(x-x_i,y-y_j)$ denotes the peak concentration distribution function for this peak centered at $(x_i,y_j)$.

From a modeling perspective, the peak concentration distribution $\psi_{ij}(x-x_i,y-y_j)$ roughly follows a Gaussian or bell-shaped distribution, with peak summit of $p_{ij}$ located at $(x_i,y_j)$. However, the exact model of each peak profile and the joint peak distribution in Equation (\ref{model}) is challenging, sometimes impossible, to determine effectively for three related and compelling reasons:
\begin{itemize}
\item[(i)] Experimental variability introduces baseline uncertainty, which changes non-linearly across time, and thus, renders an effective model for $\psi(\cdot)$ difficult to frame across the entire time period of the experimental run;
\item[(ii)] Co-eluting peaks, i.e., peaks with overlapping distribution functions are challenging to deconvolve;
\item[(iii)] Precise modeling of peak profiles is impractical, if not impossible, across hundreds of hydrocarbons, where each intricate joint peak distribution can uniquely fingerprint thousands of closely related oil reservoirs in a petroleum-rich area (Gulf of Mexico alone has $\sim 3000$ active oil sources within geographic proximity of the \emph{Deepwater Horizon} spill.
\end{itemize}

Therefore, adopting a model-based approach to petroleum forensics is rife with statistical assumptions that are impossible to assert across a broad variety of test cases. This motivates to adopt a method that is not explicitly dependent on an underlying peak distribution model, but rather, exploit the rich diversity of the peak profile.  We aim to prioritize on major peaks that dominate the peak topography, while accounting for the minor peaks that are individually less significant, but cumulatively contribute to the overall distribution. A recent attempt to achieve this using peak manifolds defined across clusters of GC$\times$GC pixels is documented in \cite{asilomar} along with data compression implications in \cite{dcc2015,dcc2016}. However, these pixel-driven techniques did not achieve robust source differentiation due to disambiguation and dimensional challenges discussed below.

\subsection{Computational and signal processing challenges to robust pattern recognition}

Despite phenomenal advances in separation technology \cite{gcxgc, gc1,gc3,gc4,gcms} the fundamental challenges in petroleum forensics are well-known source disambiguation issues in the signal processing and pattern recognition community. Specifically, the challenges may be iterated as follows. 

\begin{itemize}
\item[(i)] Closely spaced oil reservoirs share bulk of their biomarker peak distributions due to the common regional fingerprint, thus effectively creating  ``near-far'' challenge for disambiguation of weaker source-specific fingerprint against the stronger regional fingerprint.
\item[(ii)] The high dimensionality of the data limits the performance and scalability of data mining algorithms due to the ''curse of dimensionality''~\cite{curse}.
\end{itemize}

The answer to this question has been sought in two complementary directions, target compound analysis \cite{target1,target2,target3} and statistical multivariate approaches \cite{chemo1,chemo2,chemo3}. While both approaches provide insight into what the source fingerprint of a petroleum reservoir might look like, neither offer a clear path to robust fingerprinting that is cognizant of the network of biomarker compounds that constitute crude oil. This is a critical and compelling gap to fill as closely spaced petroleum sources in an oil-rich locale, e.g. the Gulf of Mexico, often share regional commonalities, which translates to significant similarities in the biomarker distribution. From a pattern recognition perspective, This poses a significant problem where the regional fingerprint masks the source-specific fingerprint.

\section{Problem Statement}\label{ps}

Consider a source library $D$, with $K$ $GC\times GC$ template images, $D = \{I_1,I_2,\dots,I_K\}$ where each image represents the unique $GC \times GC$ biomarker fingerprint of an oil source. The geographical regions of these $GC\times GC$ images are saved in a set $R=\{r_1,r_2,\dots,r_K\}$ where the $i^{th}$ element in $R$ indicates the geographical region of $i^{th}$  image of $D$. We assume $K$ to be large enough that $D$ has the requisite data diversity to robustly differentiate source fingerprints, despite significant overlap between neighboring sources. The petroleum  fingerprinting problem then reduces to matching the GC$\times$GC peak distribution $\Phi_{test}(x,y)$  of a newly-extracted unknown sample, based on the observed GC$\times$GC image {$I_{test}$, to the closest source template image $I_k \in D$. Mathematically, the closest match $I_{opt} \in D$ for $I_{test}$ is determined using the similarity criterion $\mathcal{S}$ such that:

\begin{equation}\label{pssax}  
I_{opt} = \underset{1 \le k \le K}{\operatorname{argmax ~ \mathcal{S}(I_{k},I_{test})}}.
\end{equation}
 The key contribution of this work is designing this similarity criterion such that it is robust to statistical uncertainties in the peak profile distribution. We define the similarity criterion $\mathcal{S}(\cdot)$ in Section (\ref{sim}) based on the relative peak distribution of the reference and test samples. 
 
\section{Technical Approach}\label{ta}

We present the solution to the problem stated in Section \ref{ps} with the framework shown in Figure \ref{ntw}. It is noteworthy that the computational framework in Figure \ref{ntw}, though similar in architecture to an artificial neural network (ANN) \cite{ann}, it is not implemented as such, but rather employed to identify peak profile features for source fingerprinting in an unsupervised setup. The proposed framework consists of the three layers: (i) Input, (ii) Comparison and (iii) Decision. 

\begin{figure}
\centering
\includegraphics[scale=0.4]{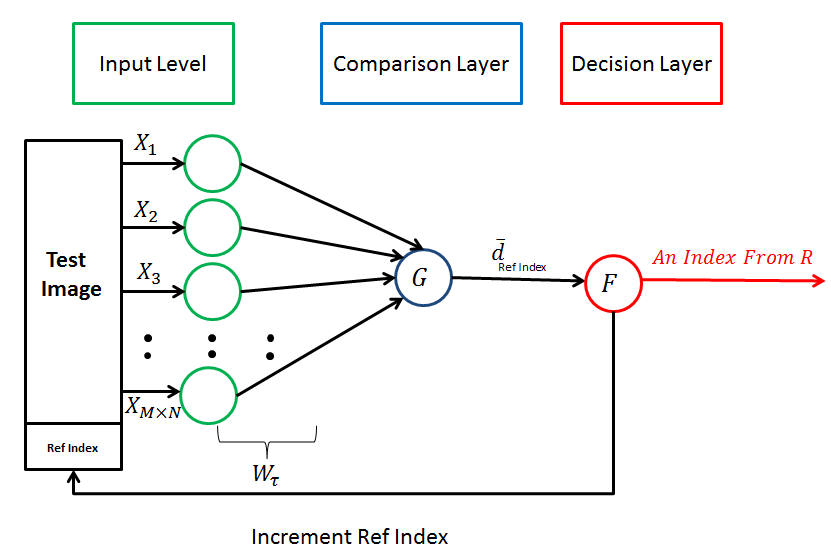}
\caption{Illustrative model of the proposed framework.}
\label{ntw}
\end{figure}

We now provide some details on the different parts of the framework in Figure \ref{ntw}. The box on the left, has two parts, first the $GC\times GC$ image of the injection under the test $(I_{test})$ and a block for the index of the reference injections from the library $D$ to be tested against $I_{test}$, which this index is initialized to one (RefIndex=1). The test image is an $M\times N$ matrix, so in the input layer we have $M\times N$ inputs, shown as $X_1$ to $X_{M\times N}$ where each of these inputs indicates one pixel of the test image. These inputs are transmitted by the weights  $W_\tau$ to the Comparison layer. In Section \ref{weights} we will discuss on the procedure to construct $W_\tau$. 

In the Comparison layer the $G$ function is implemented between $I_{test}$ and the reference image. Finally, in the Decision layer, the function $F$ has the role to determine the geographical region of $I_{test}$. $F$ has $K$ memories. First, the Reference Index is one, so $I_{test}$ is compared against the first element of $D$ and the corresponding $\bar{d}_{RefIndex}$ is saved in one memory of $F$. Note that $\bar{d}_{RefIndex}$ is a criterion of deviation of the test image from the reference image chosen from $D$. Once this comparison is done completely and $\bar{d}_{RefIndex}$ is saved, $F$ sends an \enquote{Increment Reference Index} command to  the leftmost box and in the next time the test image should be compared against the second element of $D$ and so on. 
Finally once Reference Index exceeds the size of the dictionary, $K$, $F$ outputs the index having the lowest $\bar{d}_{RefIndex}$ which is the index of the geographical region of $I_{test}$ from the set $R$. 

    

   

\subsection{Similarity Criterion}
\label{sim}
As discussed comprehensively in \cite{asilomar,ccj} comparing two $GC\times GC$ images is the comparison between their corresponding peaks at the same location of their image.

We sweep the image by going along each of its $N$ columns and save the local maxima, detected along each of the columns as peaks. Suppose we are to compare two images, $I_{ref}$ and $I_{test}$. At one location, like $(x_0,y_0)$ the amplitude of the peaks in $I_{ref}$ and $I_{test}$ are $p_{ref}$ and $p_{test}$, respectively. In \cite{asilomar,ccj} the similarity between these two peaks is defined as:

\begin{equation}\label{sim_form}
Sim(p_{ref},p_{test}) = max(\frac{p_{ref}}{p_{test}}, \frac{p_{test}}{p_{ref}})
\end{equation}

As known, the function $max(\alpha,\alpha^{-1})$ has a value greater or equal to one for all $\alpha \in \mathcal{R}$, $\alpha \neq 0$. Therefore, if the function $Sim(\cdot)$ has a value of one, then $\alpha=\alpha^{-1}$ or in our case $p_{ref}=p_{test}$. Any difference between the values of $p_{ref}$ and $p_{test}$ results in a value greater than one for $Sim(\cdot)$. We can call two peaks as similar once $Sim(\cdot)$  for their corresponding location is unity. However, to account for experimental variability during the process of producing the $GC\times GC$ image, we relax the definition of similarity to within an uncertainty bound $\epsilon$. Hence, we define a peak-ratio parameter $\tau$ as:

\begin{equation} \label{epsil} 
\tau = 1+\epsilon  ~    (\epsilon>0) 	 
\end{equation}

We claim two peaks as similar, once the $Sim(\cdot)$ function for their corresponding location is less than or equal to $\tau$ $(Sim(\cdot) \le \tau)$. The parameter $\epsilon$ indicates the amount of deviation that is acceptable for us to consider two peaks as similar; the more the value of $\epsilon$ is, the less strict we are in the definition of similarity between the two peaks. In an extreme case, if we set $\epsilon=\infty$, all of the peaks will be considered as similar because $Sim(\cdot)$ is always less than $\infty$.

\subsection{How to set $W_\tau$}\label{weights}

There are $M\times N$ inputs, $X_1$ to $X_{M\times N}$, but as discussed in the previous section, just the local maxima of the image are saved and used as the comparison. Therefor, the weights for the non-peak pixels or inputs are set to zero. For calculating the weight of one peak like $p_{test}$ at the location $(x_0,y_0)$, once the peak in the same location in the reference image is $p_{ref}$, while the peak-ratio is $\tau$, we perform the following two-stage process:
\vspace{3 mm}

\begin{itemize}
 
\item[\textbf{1.}]  Calculate the $Sim(\cdot)$ for the location of the peaks as in Equation \ref{sim_form}.
\item[\textbf{2.}]  

   \begin{itemize}
   \item[$a.$] If $Sim(\cdot) \le \tau$ then replace both $p_{ref}$ and $p_{test}$ by a common peak, $p_{common}$:
                    
                    \begin{equation*}
                     	p_{common}=min(p_{ref},p_{test}).
                    \end{equation*}
    so the weight of $p_{test}$ will be:
    
                     \begin{equation*}
                     	w_{test}(x_0,y_0) = \frac{p_{common}}{p_{test}} .
                    \end{equation*}
                    
    \item[$b.$] Otherwise, keep the exact values of $p_{ref}$ and $p_{test}$ which means $w_{test}(x_0,y_0)=1$.
     \end{itemize}

\end{itemize}
   
The reason which we replace $p_{ref}$ and $p_{test}$ by $p_{common}$ is that, we have assumed once $Sim(\cdot)$ is less than equal to $\tau$ for that location, these two peaks are similar, and therefore we manually assign an equal value to them. We use the minimum value as the common peak for both of the reference and test images, because the minimum peak amplitude is common between the two peaks.

\subsection{How to set $\epsilon$} \label{howtoset} 

Suppose we have the library $R=\{I_1^{k_1}  ,I_2^{k_2}  ,\dots,I_K^{k_N} \}$, where the $i^{th}$ element of the library, $I_i^{k_i}$, or $i^{th}$ family,  means we have $k_i$ number of $GC\times GC$ images for the region indexed by $i$. For each of these image families, we also learn an $\epsilon(i)$ with some training members of the family. For each of these training set injections, let's say $I_{train}$, we first construct its maximum-along-Interval representation as following:

\begin{itemize}
 
\vspace{3 mm}
\item[\textbf{1.}]  We pass each column of $I_{train}$ to the proposed network and apply its corresponding $W_\tau$ with respect to a reference image of the family.

\item[\textbf{2.}]  For each column of $I_{train}$, named as $I_{train}^{c}$, which is a time series, we do the following:

\begin{equation}\label{max}  
c_{max}(i) = \underset{i \in \{1,2, \dots, \frac{n}{w}\}}{\operatorname{argmax ~ \textit{$I_{train}^{c}(i:i+w)$}}}.
\end{equation}

\end{itemize}

In other words, for each column we have $\frac{n}{w}$ intervals, and we replace each interval with its maximum value. Therefore, we have constructed the maximum-along-Interval representation of $I_{train}$. Once, we have done the above-mentioned two stages for all of the images in the training set, we compute the average  $L_2$ norm distance of the all of the images in the training set. This average $L_2$ norm distance is equivalent to the deviation of the training images after applying the proposed two stage method. Then we sweep over an increasing range of $\epsilon$'s and compute its corresponding deviation. It is noteworthy that the amount of the deviation is disproportional to the value of $\epsilon$, i.e, as we increase $\epsilon$,  the deviation among the images decrease. This is because, as denoted in equation \ref{epsil}, the more the value of $\epsilon$, the more similar two arbitrary peaks are to each other and the less strict we are in our definition of similarity. 

This can be seen in Figure \ref{tau}, where the deviation is decreasing by increasing $\epsilon$, as we are to set the $\epsilon$ for the family of images from Macondo well. In this figure, we have set the first seven Macondo injections as the training set for calculating the $\epsilon$. But we wish to set the value of $\epsilon$ not too far from zero, since, at least in an extreme case, setting $\epsilon = \infty$ would indicate any two peaks with different values as being similar. So we wish to choose a good choice for $\epsilon$ where the deviation is low and it is in its minimum possible value. As can be seen in Figure \ref{epsil}, the deviation drops until $\epsilon = 0.5$ but remains the same for higher values. So the best choice for $\epsilon$ would be half.

\begin{figure*}[t]
    \centering
    \includegraphics[scale=0.35]{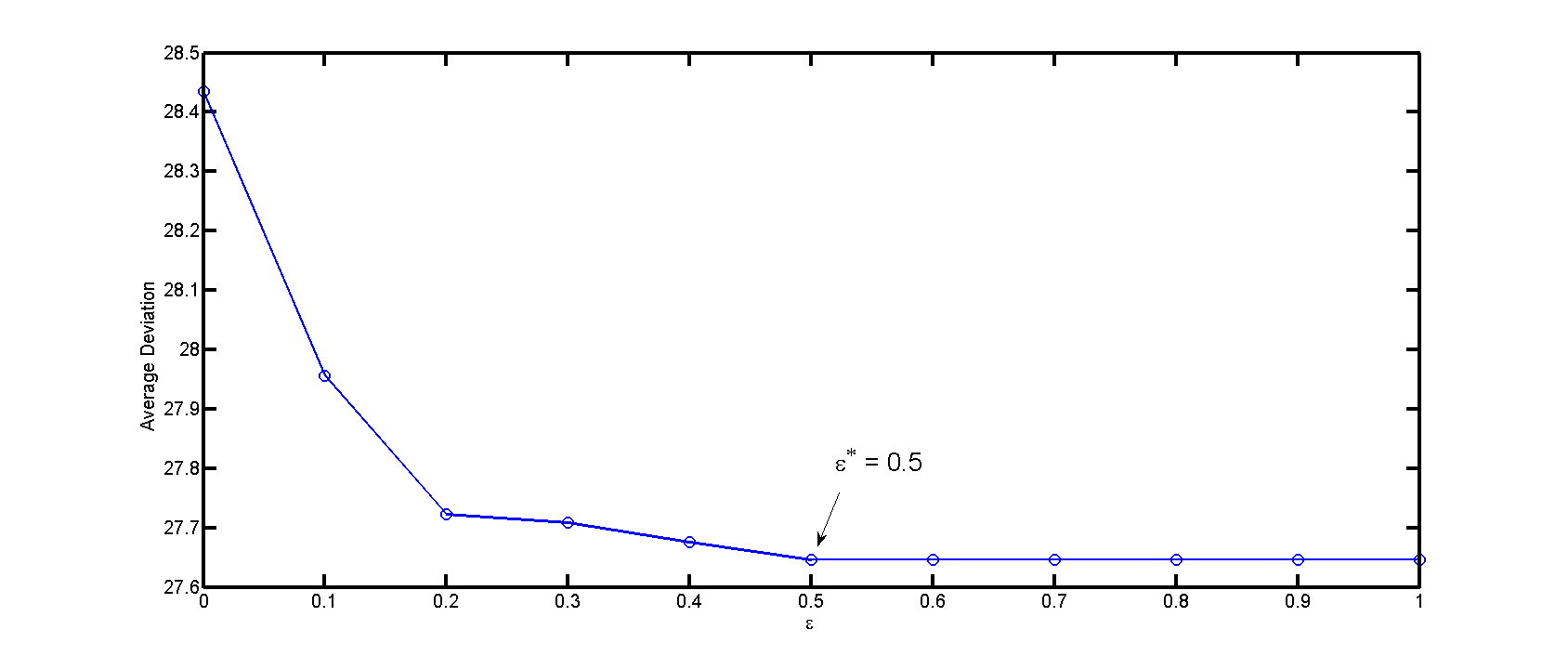}
    \caption{Optimal choice of $\epsilon$}
    \label{tau}
\end{figure*}

\section{Function $G$}
As discussed earlier, in our proposed framework, the Comparison layer which is implemented by function $G$, computes the distance between the images under the test and the reference image using their SAX representations. In the following, we briefly touch upon the SAX method.
\subsection{SAX algorithm} \label{sax_algo}

The comprehensive study on Symbolic Aggregate ApproXimation (SAX) algorithm can be found in \cite{2,3,4}. Here, we touch upon it briefly, as we will use it in our proposed framework in Figure \ref{ntw} as function $G$ in the Comparison layer. Formally, SAX is a method to represent time series of size $n$ using a string of arbitrary size $w (w<n)$. Applying SAX, one can control the dimensionality reduction imperative when dealing with large time series. As formally defined in \cite{2}, a time series $C =c_1c_2c_3 \dots c_n$ of length $n$ can be represented in a $w$-dimensional space by a vector $\bar{C} =\bar{c_1} \bar{c_2} \bar{c_3} \dots \bar{c_w}$. The $i^{th}$  element of $C$ is calculated by the following equation:

\begin{equation}
\bar{c_i} = \frac{w}{n}\times \sum^{j=\frac{n}{w}i+1}_{j=\frac{n}{w}(i-1)+1} c_j   	
\end{equation}

In other words, the time series is divided into $w$ intervals of the same size, then the data in each of these intervals is replaced by the mean of the data. This representation is called the Piecewise Aggregate Approximation (PAA) representation of the time series. Now each of these intervals is mapped into one symbol. These samples are chosen out of some set with the cardinality of $\alpha$.
After applying the SAX method, the final symbol representation of the times series $C$ will be:


\begin{equation}
\hat{C} = \hat{c_1}\hat{c_1} \dots \hat{c_w}     
\end{equation}
Note that $C$ is the original times series, $\bar{C}$ is its PAA representation and $\hat{C}$ is its SAX representation.

\begin{figure*}
    \centering
    \includegraphics[scale=0.4]{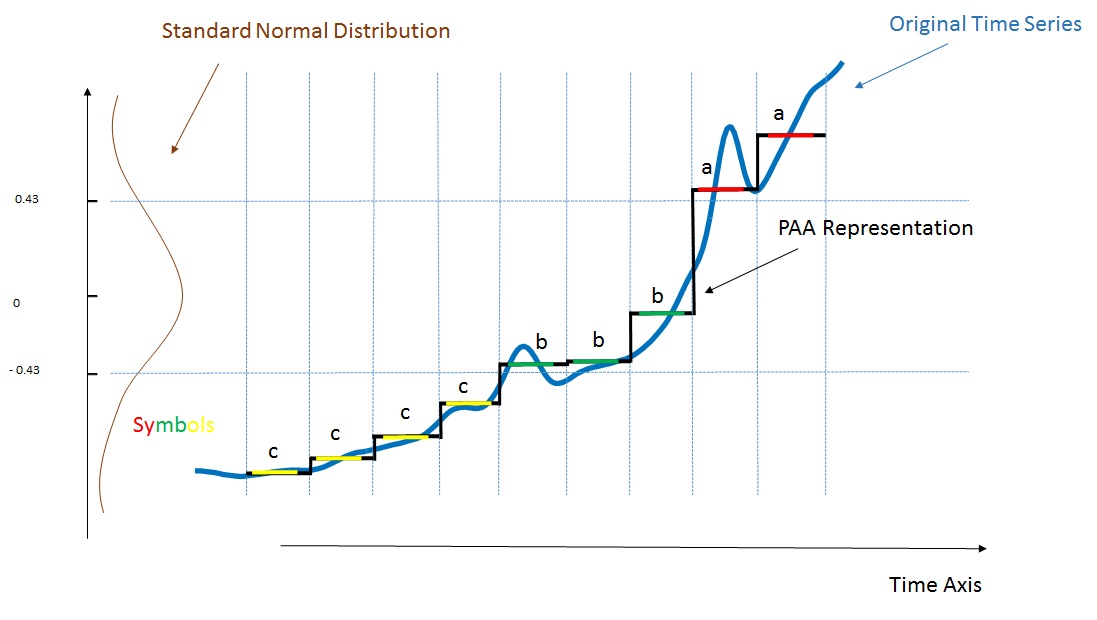}
    \caption{The PAA and SAX representation of a model time series. In this figure, the there are three symbols, $a$,$b$ and $c$. The time axis has been sliced into nine intervals. The SAX representation of the time series in this case would be  $\hat{C}=ccccbbbaa$.}
    \label{sax}
\end{figure*}

\subsection{Calculating the distance between two time series using their PAA and SAX representations}


 
 Suppose we have two time series $C_1$ and $C_2$. The Euclidean distance between the PAA representation of the two time series is :
 
 \begin{equation}
     \bar{d}=\sqrt{\sum^{w}_{i=1}(\bar{c}^i_1 - \bar{c}^i_2)^2}
 \end{equation}

 Where $\bar{c}_1^i$  and $\bar{c}_2^i$ are the $i^{th}$ element of the $\bar{C}_1$ and $\bar{C}_2$, respectively.
  After constructing the symbol representations of $C_1$ and $C_2$, we wish to calculate the symbol distance of their representations. Therefore, we need a look-up table in order to have the distance between the symbols. Such a table is given in Table \ref{SAX symbol distance}. Then, The Euclidean distance between the two SAX representation of $C_1$ and $C_2$ is given as \cite{2}:

 \begin{equation}
 \hat{d}=\sqrt{\frac{n}{w}\sum^{w}_{i=1}[dist(\hat{c}^i_1 - \hat{c}^i_2)]^2}
 \end{equation}
 
 Where $\hat{c}_1^i$  and $\hat{c}_2^i$ are the $i^{th}$ element of the $\hat{C}_1$ and $\hat{C}_2$, respectively.

 \begin{table*}[h!p]
\caption{Look-up table for the distance  of SAX symbols}
\label{SAX symbol distance}
\centering
    \begin{tabular}{ | l | l | l | l | l | l | l | l | l | l | l |}
    \hline
    Alphabet & a & b & c & d & e & f & g & h & i & j \\ \hline
    a & 0 & 0 &0.1936& 0.5776& 1.0609& 1.6384&2.3409& 3.2400 & 4.4944 &6.5536  \\ \hline
    b & 0 & 0 & 0&0.1024&0.3481&0.7056&1.1881&1.8496&2.8224&4.4944   \\ \hline
    c & 0.1936 & 0 & 0&0& 0.0729&0.2704&0.5929&1.0816&1.8496&3.2400 \\ \hline
    d & 0.5776 & 0.1024 &0&0&0&0.0625&0.2500&0.5929&1.1881&2.3409  \\ \hline
    e & 1.0609 & 0.3481 &0.0729&0&0&0&0.0625&0.2704&0.7056&1.6384    \\ \hline
    f & 1.6384 & 0.7056 & 0.2704&0.0625&0&0&0&0.0729&0.3481&1.0609 \\ \hline
    g & 2.3409 & 1.1881 &0.5929&0.2500&0.0625&0&0&0&0.1024&0.5776  \\ \hline
    h & 3.2400& 1.8496 & 1.0816 &  0.5929&0.2704&0.0729&0&0&0&0.1936 \\ \hline
    i & 4.4944 & 2.8224 & 1.8496 & 1.1881&0.7056&0.3481&0.1024&0&0&0 \\ \hline
    j & 6.5536 & 4.4944 & 3.2400 & 2.3409& 1.6384&1.0609&0.5776&0.1936&0&0\\ \hline

    \end{tabular}

\end{table*}

 
$L_2$ norm distance (Euclidean distance) between $C_1$ and $C_2$ is:
\begin{equation}
 d=\sqrt{\sum^{n}_{i=1}(c^i_1 - c^i_2 )^2}         	            
 \end{equation}
 
The two-dimensional correlation between the two images $I_{ref}$ and $I_{test}$ images are computed as following:

\begin{multline}
    Corr2(I_{ref},I_{test}) = \\ \hfill
    \frac{\sum_m\sum_n[ (I_{ref}^{m,n} - \bar{I}_{ref}) \times (I_{test}^{m,n} - \bar{I}_{test})]}{\sqrt{\sum_m\sum_n (I_{ref}^{m,n} - \bar{I}_{ref})^2 \times \sum_m\sum_n (I_{test}^{m,n} - \bar{I}_{test})^2 }}
\end{multline}
    
where $m$ and $n$ represent the first and the second dimension of the image, respectively. $\bar{I}_{ref}$ and $\bar{I}_{test}$ also represent the mean value of the reference and test images, respectively. 
We applied PCA using two principle components and the similarity between $I_{ref}$ and $I_{test}$ is calculated as the correlation between the scores in their corresponding principle components. PTM score is also completely explained in \cite{patent, asilomar}.

Basically, we quantized the image treated as a time series using SAX symbols and then apply the similarity criterion in PTM, therefore we call our proposed network or method, Quantized Peak Topography Map or briefly QPTM.

In order to evaluate QPTM, we test it against the other established comparison criteria, $L_2$ norm distance, Correlation, PCA, PTM score and SAX. Note that in these methods, we don't use the proposed network, we just apply these methods directly and use them as the difference criteria. Specifically, we have used SAX algorithm both individually as the comparison criterion, which is named "SAX" method in the figures and also as the Comparison layer in function $G$ in QPTM. Also in the method named "PAA" in the figures, we have applied the dataset into the proposed network but in the Comparison layer, we just computed the Euclidean distance between the PAA representations of the reference and test images.

  \section{Result} \label{result}
 \subsection{Data description}
 
 We verify the proposed method in this work with a dataset of thirty-four injections with $GC\times GC$ patterns from different parts of the world. Of particular interest, there are seventeen injections from the Gulf of Mexico from which fourteen of them are from Macondo well, one injection from Eugene Island, one injection from Southern Louisiana and one injection is a natural seep in the Gulf of Mexico. There also exist three standard NIST injections unrelated to the Gulf of Mexico in our model dataset. The rest fourteen injections are from the other parts of the world.
 
 \begin{figure*}[h]
 \centering
 \includegraphics[scale=0.3]{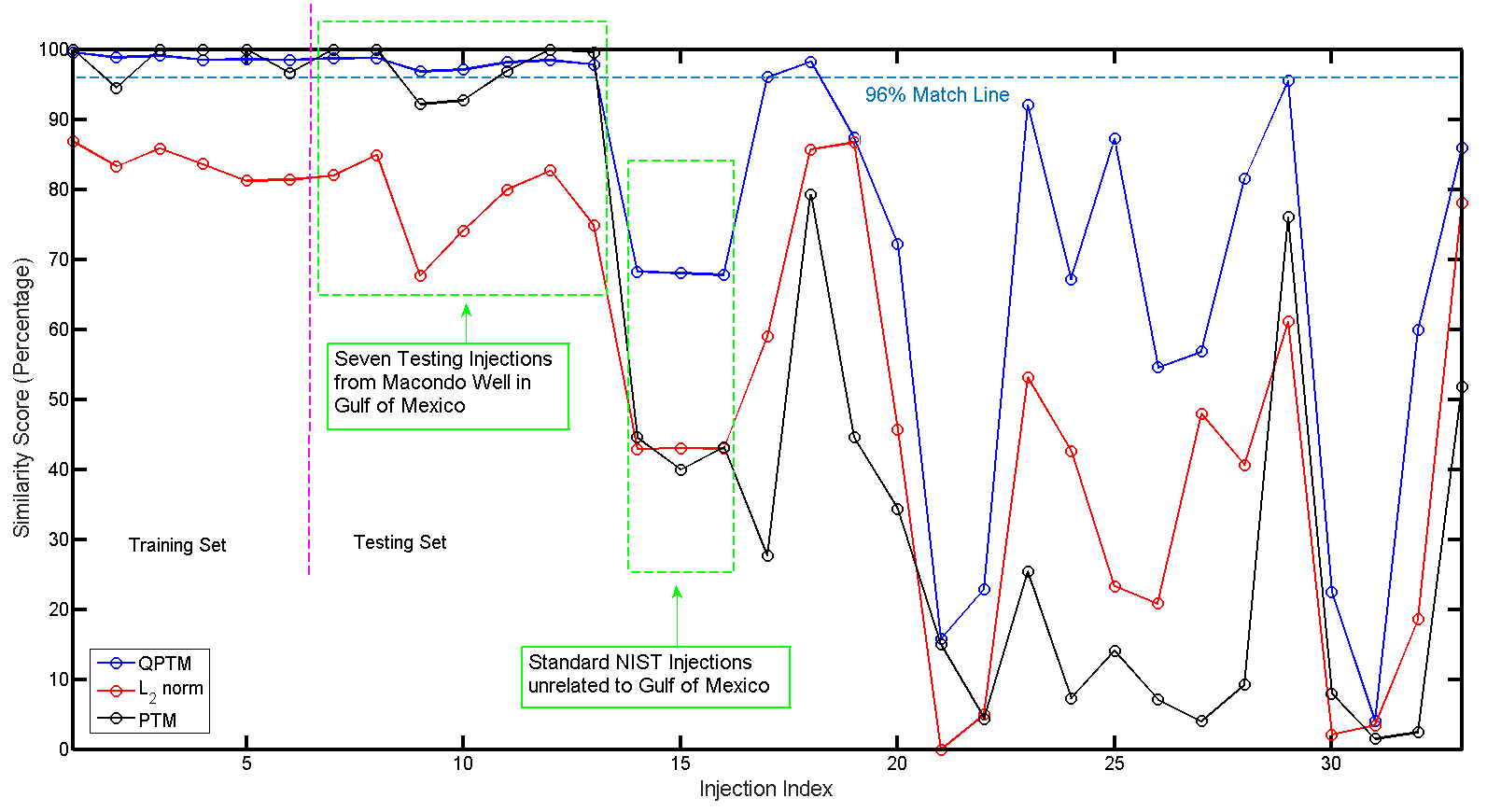}
 \caption{The percentage of similarity of thirty three samples from the model dataset to the first reference sample from Macondo well (1). }
 \label{r1}
 \end{figure*}

  \begin{figure*}[h]
 \centering
 \includegraphics[scale=0.3]{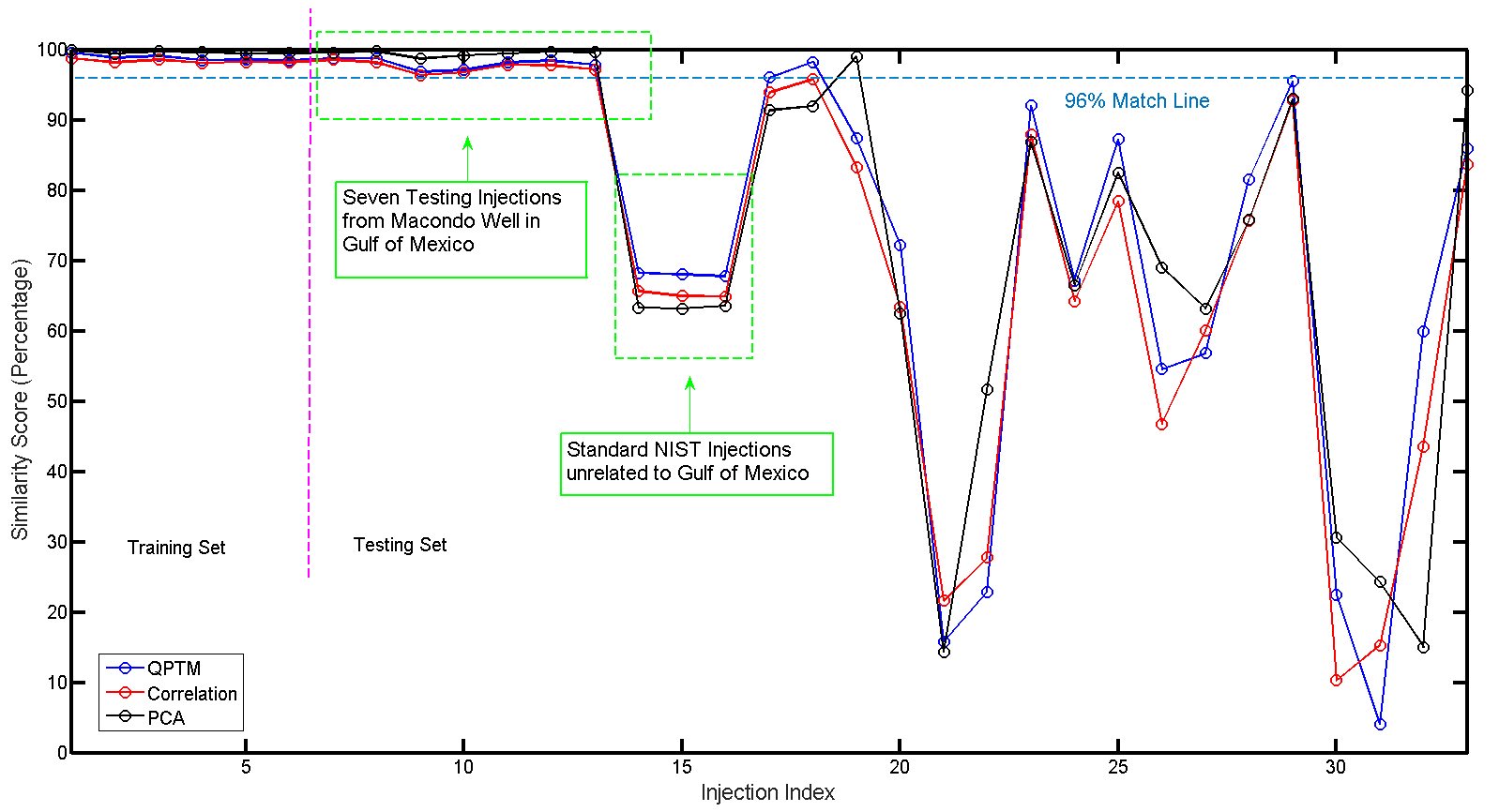}
 \caption{The percentage of similarity of thirty three samples from the model dataset to the first reference sample from Macondo well (2). }
 \label{r2}
 \end{figure*}

  \begin{figure*}[h]
 \centering
 \includegraphics[scale=0.3]{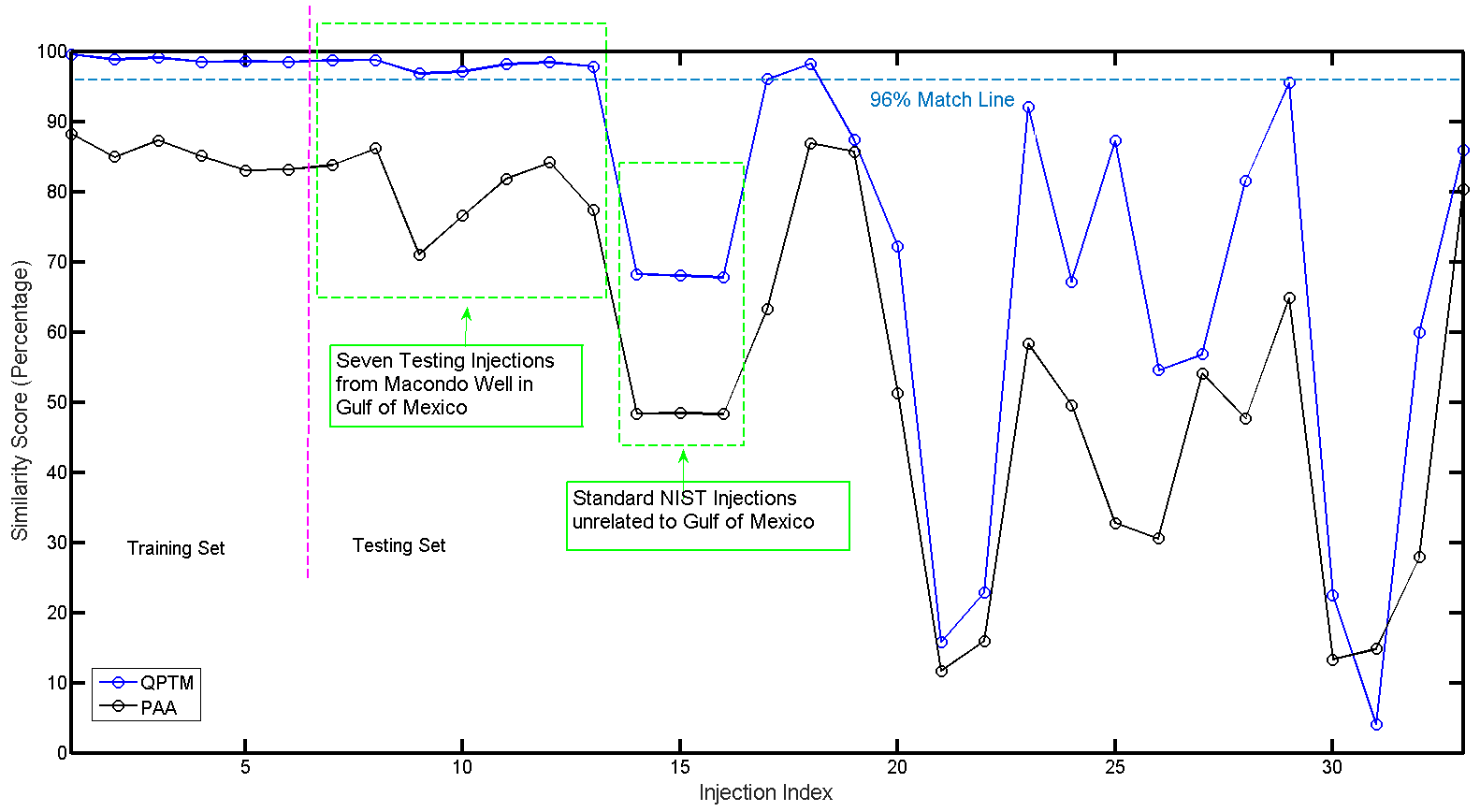}
 \caption{The percentage of similarity of thirty three samples from the model dataset to the first reference sample from Macondo well (3). }
 \label{r3}
 \end{figure*}

 \begin{figure*}[h]
 \centering
 \includegraphics[scale=0.3]{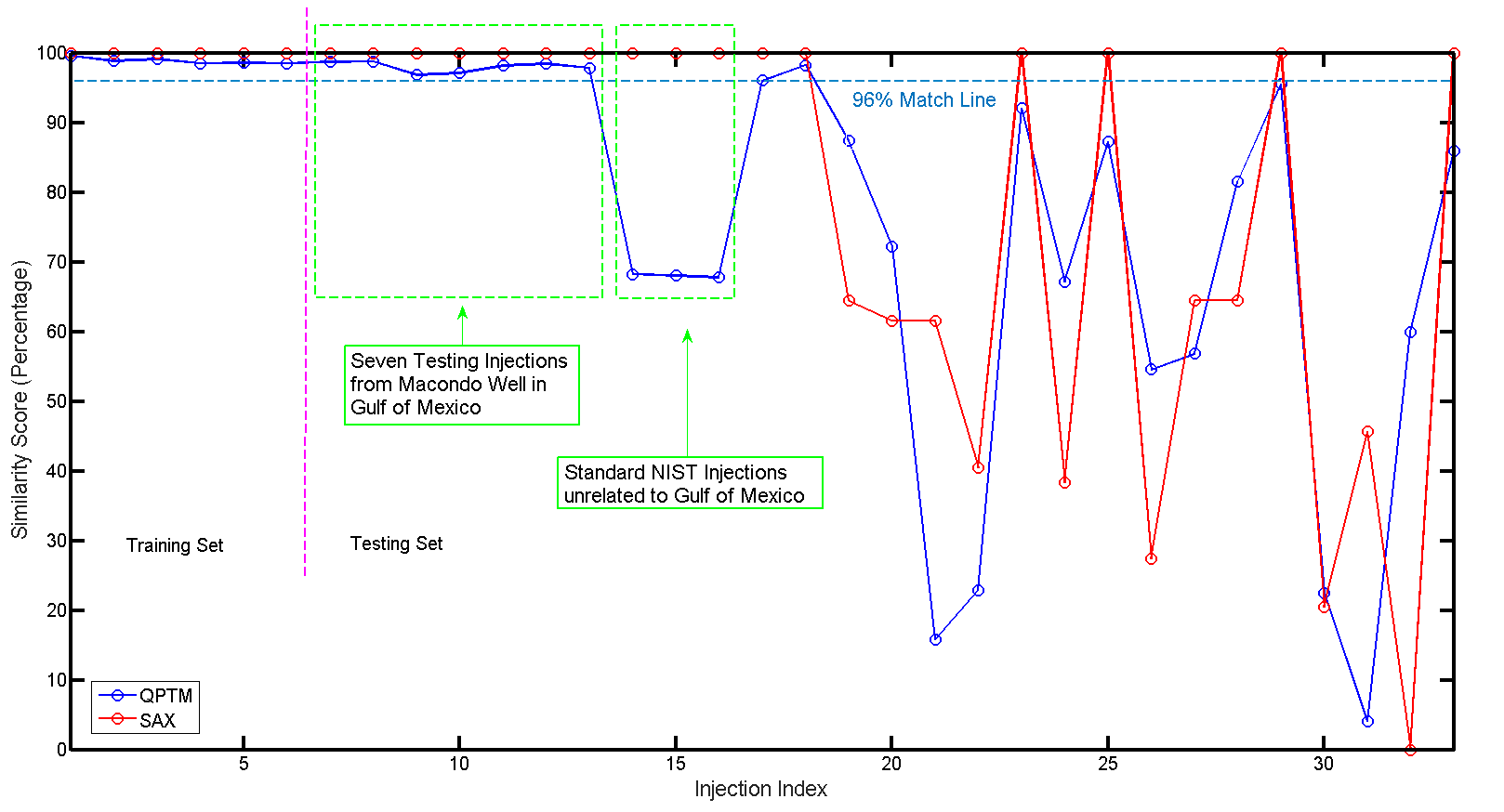}
 \caption{The percentage of similarity of thirty three samples from the model dataset to the first reference sample from Macondo well (4). }
 \label{r4}
 \end{figure*}

  \begin{figure*}[h]
 \centering
 \includegraphics[scale=0.35]{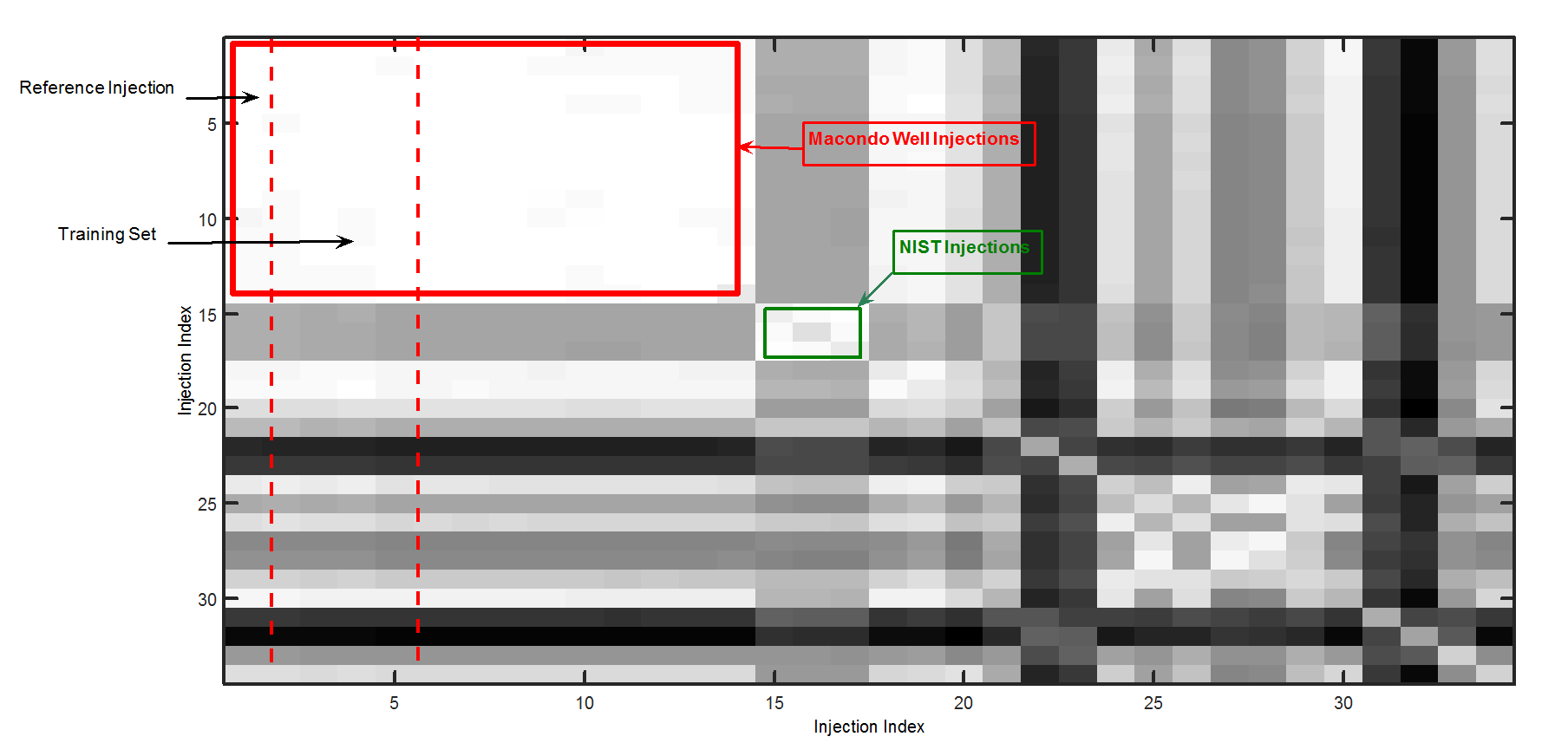}
 \caption{Cross-QPTM score for the thirty-four injections using QPTM for $\epsilon = 0.5$. The white box on upper left are injections from the Macondo well which have a very high percentage of similarity. Note that, in each row, we set the first injection as the reference sample, the rest six injections as the training injections to evaluate the $\epsilon$ and then the rest would be the testing set. NIST injections are also shown at the middle of the image. }
 \label{34by34}
 \end{figure*}



As stated in \ref{ps} for a given test image, we should find the most similar image from the library in which the similarity is also above a threshold. But it requires to have at least one image from each of the geographical region, in which we don't have in our model dataset. Therefore, we just consider the case where we have twenty-seven test samples and test them against the first element of $D$, which is an injection from Macondo well. Note that our model dataset has thirty-four injections, but because we have set aside seven of the injections from Macondo well as the training set for computing $\epsilon$, we end up having a dataset of size twenty-seven.
Once, we compare our test images against this Macondo injection, the question will turn into the detection of the Macondo samples from the test images. In other words, we are to detect either each of the test images is from Macondo well or not.
 For better illustrating the comparison between QPTM and the other methods, we have shown the results of comparison of our twenty-seven injection with respect to the reference injection from the Macondo well (the first element of $D$) in four consecutive figures, Figure \ref{r1}, \ref{r2}, \ref{r3} and \ref{r4}, in each of them comparing QPTM against some of the other methods.
 
 \subsection{Discussion over the results} \label{discussion}
 In each of these figures, we are testing the other injections against the first injection in our dataset which is from Macondo well. We have used the reference injection along with six other injections as the training set for learning the parameter $\epsilon$ for the family of the images from Macondo well, which has been explained comprehensively in Section \ref{howtoset}.

 As can be seen, in all of the Figures \ref{r1} to \ref{r4}, the first thirteen injections have a high percentage of match with respect to the reference injection. The first six injections are the Macondo injections used for learning $\epsilon$ of Macondo well and the rest seven injections are some other injections from the Macondo well. Therefore, these figures show that the proposed method have performed well in detecting the Macondo injections. In these figures, we have also shown the standard NIST injections. These injections have almost exactly similar percentage of match to the reference injection. In Figure \ref{34by34} we have also shown the cross QPTM score for the thirty-four injections using $\epsilon=0.5$.
 
The problem statement is detecting the injections from different geographical regions and classifying them accordingly. Therefore, our proposed network serves as a classifier. The commonly-used metric to observe the accuracy of a classifier is its confusion matrix. In table \ref{confusionMatrix} we have shown the confusion matrix for detecting the injections from Macondo well, using different methods in matrix $C$. $C_{i,j}$ element of $C$ represents the number of the injections being from region indexed by $i$ and being detected as $j$. In our case the confusion matrix is a two by two matrix, as our classifier has two classes, being from Macondo or not. The first and second rows of $C$ correspond to the injections being from Macondo well and some other source rather than Macondo, respectively. Likewise, the first and second column of $C$ represents the injections being detected as Macondo well and some other source, respectively.

As can be seen in the figures \ref{r1} to \ref{r4}, we have chosen $\theta  = 96 \%$ as the classification threshold. In Table \ref{accpre} we have shown the values for Accuracy, precision, Sensitivity, Specificity and $F_1$ scores for the different methods. 

As can be seen QPTM has the highest scores in Accuracy, Sensitivity and $F_1$ among the others. In Table \ref{mean} we have computed the mean percentage of match between all Maconod injections against Macondo, Eugene Island (EI), Southern Louisiana Crude (SLC) and the natural Seep (NS) from Gulf of Mexico. 
The best method for detecting the Macondo injections is the one which has the highest percentage of match in the first column and has the lowest in the other three columns. Remember, the other three injections, EI, SLC and NS are closely located to the injections from the Macondo well as they all are from the Gulf of Mexico. SAX algorithm could not distinguish between Macondo injections and the other three. QPTM and PCA have the highest percentage of match for the injections from the Macondo well. PTM has performed better than the others in rejecting the other injections, as the percentage of match of the other three injections to the Macondo well injections is lower than the others.

\begin{table*}
\centering

\caption{Confusion matrix for QPTM -  SAX - PTM - Correlation - PAA - $L_2$ norm - PCA}\label{confusionMatrix}

\begin{tabular}{||c | c ||} 

 \hline
 Method & Confusion Matrix \\ [0.5ex] 
 \hline\hline

QPTM & $ C  = \begin{bmatrix}
    7  & 0 \\
    2  & 18 \\
    \end{bmatrix}$  
    
  \\

SAX & $C = \begin{bmatrix}
    7  & 0 \\
    9  & 11 \\
    \end{bmatrix} $

\\

PTM & $C =  \begin{bmatrix}
    5  & 2 \\
    0  & 20 \\
    \end{bmatrix} $

\\

Correlation & $C = \begin{bmatrix}
    7  & 0 \\
    3  & 17 \\
    \end{bmatrix} $

\\ 

PAA &  $C  = \begin{bmatrix}
    0  & 7 \\
    0  & 20 \\
    \end{bmatrix} $
    
\\

$L_2$ norm & $C = \begin{bmatrix}
    0  & 7 \\
    0  & 20 \\
    \end{bmatrix} $
    
\\

PCA & $C = \begin{bmatrix}
    7  & 0 \\
    4  & 16 \\
    \end{bmatrix}  $
    
\\

 \hline
\end{tabular}

\end{table*}

\begin{table*}
\centering
\caption{Accuracy, Precision, Sensitivity, Specificity and $F_1$ score calculation for different methods}\label{accpre}

\begin{tabular}{||c | c | c | c | c | c||} 

 \hline
 Metric & Accuracy & Precision & Sensitivity & Specificity & $F_1$ score\\ [0.5ex] 
 \hline\hline

QPTM & \textbf{0.9259} &  0.7778 & \textbf{1} &  0.9 & \textbf{0.8750} \\  

SAX & 0.67 &  0.4375 & \textbf{1} &  0.55 & 0.608 \\

PTM &  0.9259 &  \textbf{1} & 0.7143 &  \textbf{1} & 0.8333 \\ 

Correlation & 0.8889 &  0.7  & \textbf{1} & 0.85 &  0.8235\\

PAA &  0.7407 &  NaN  &  0 & 1 &    0\\

 $L_2$ norm & 0.7407 & NaN  & 0 & \textbf{1} &  0\\ 
 
 PCA & 0.7407 & NaN  & 0 & \textbf{1} &  0\\


 \hline
\end{tabular}

\end{table*}

 \begin{table*}[h]
\centering

\caption{Percentage match (Mean $\pm $ standard deviation) between different Gulf of Mexico sources against Macondo injections for QPTM -  SAX - PTM - Correlation - PAA - $L_2$ norm - PCA}\label{mean}
\begin{tabular}{||c |c| c| c| c||} 
 \hline
 Method & Macondo vs. Macondo & Eugene Island vs. Macondo & Southern Louisiana Crude (SLC) vs. Macondo &  Natural seep vs. Macondo \\ [0.5ex] 
 \hline\hline
 
 SAX & $ 100\%$ & $100\%$ & $ 100\%$ & $   100\%$ \\ 
 
 QPTM & $  99.1487\% \pm  0.5743\%$ & $ 96.1997\% \pm   0.4973\%$ & $  95.3152\% \pm  0.5544\%$ & $   87.5703\% \pm  0.2041\%$ \\ 
 
  PTM & $ 97.7816\% \pm   3.5909\%$ & $ 79.3793\% \pm  5.3834\%$ & $ 36.4339\% \pm  10.1465\%$ & $ 45.4172\% \pm   2.5952\%$ \\

 Correlation & $98.7046\% \pm .8238\%$ & $94.0824\% \pm .8173\%$ & $92.3692\% \pm 1.1816\%$& $83.3948\% \pm .24\%$  \\ [1ex] 

 PAA & $ 92.0109\% \pm     4.1716\%$ & $ 86.2351\% \pm      2.5682\%$ & $  79.5171\% \pm   5.0826\%$ & $  72.9431\% \pm   3.9146\%$ \\ [1ex] 

 $L_2$ norm  & $92.5275\% \pm 3.9548\%$ & $87.7106\% \pm 2.5333\%$ & $81.3793\% \pm 5.1144\%$& $75.0651\% \pm 4.5358\%$  \\ [1ex] 

 PCA & $99.8313\% \pm 0.1631\%$ & $91.9702\% \pm .1025\%$ & $91.8002\% \pm .0926\%$ & $97.8385\% \pm .5136\%$ \\

 \hline
\end{tabular}

\label{table:1}
\end{table*}

\section{Conclusion}
\label{conclusion}
Source differentiation in petroleum fingerprinting using hydrocarbon biomarkers is a classic "near-far" disambiguation challenge from the signal processing perspective, where "near" and "far" represent the stronger regional fingerprint and weaker source-specific fingerprint respectively. In this paper, we propose a signal representation method called quantized Peak Topography Mapping (QPTM) along with related differentiation techniques aimed at better separation between the petroleum injections. The proposed method provides a layered interpretation of two-dimensional gas chromatography images. The main contributions of this work is interpreting intricate peak profile distributions as a multi-dimensional time series, which is quantized resulting in its symbolic aggregate representation. Thus, the distance between two time series is reduced to the difference between their symbolic representations, which is performed using a look-up table (Table \ref{SAX symbol distance}). QPTM out-performed the currently established methods for better classification of the petroleum injections in terms of accuracy, sensitivity and $F_1$ score. 

\section*{Acknowledgment}
This research was made possible in part by a grant from the University of Iowa, DEEP-Consortium and WHOI interdisciplinary study award.
\bibliographystyle{IEEEtran}

\bibliography{refs}

\end{document}